\documentstyle[12pt]{article}
\def\Re{{\cal R \mskip-4mu \lower.1ex \hbox{\it e}\,}}
\def\Im{{\cal I \mskip-5mu \lower.1ex \hbox{\it m}\,}}

\def\beq{\begin{equation}}
\def\eeq{\end{equation}}

\def\sp6{Sp(6)_L \times U(1)_Y}

\def\sub#1{_{\lower.25ex\hbox{$\scriptstyle#1$}}}
\def\sul#1{_{\kern-.1em#1}}
\def\sll#1{_{\kern-.2em#1}}
\def\sbl#1{_{\kern-.1em\lower.25ex\hbox{$\scriptstyle#1$}}}
\def\ssb#1{_{\lower.25ex\hbox{$\scriptscriptstyle#1$}}}
\def\sbb#1{_{\lower.4ex\hbox{$\scriptstyle#1$}}}

\def\GeV{\,{\rm GeV}}

\def\JL{J. L. Lopez}
\def\DVN{D. V. Nanopoulos}

\def\to{\rightarrow}

\def\mh{\ifmmode m\sbl H \else $m\sbl H$\fi}
\def\mch{\ifmmode m_{H^\pm} \else $m_{H^\pm}$\fi}
\def\mt{\ifmmode m_t\else $m_t$\fi}
\def\mc{\ifmmode m_c\else $m_c$\fi}
\def\mz{\ifmmode M_Z\else $M_Z$\fi}
\def\mw{\ifmmode M_W\else $M_W$\fi}
\def\mws{\ifmmode M_W^2 \else $M_W^2$\fi}
\def\mhs{\ifmmode m_H^2 \else $m_H^2$\fi}
\def\mzs{\ifmmode M_Z^2 \else $M_Z^2$\fi}
\def\mts{\ifmmode m_t^2 \else $m_t^2$\fi}
\def\mcs{\ifmmode m_c^2 \else $m_c^2$\fi}
\def\mchs{\ifmmode m_{H^\pm}^2 \else $m_{H^\pm}^2$\fi}
\def\ztwo{\ifmmode Z_2\else $Z_2$\fi}
\def\zone{\ifmmode Z_1\else $Z_1$\fi}
\def\mtwo{\ifmmode M_2\else $M_2$\fi}
\def\mone{\ifmmode M_1\else $M_1$\fi}
\def\tb{\ifmmode \tan\beta \else $\tan\beta$\fi}
\def\xw{\ifmmode x\sub w\else $x\sub w$\fi}
\def\ch{\ifmmode H^\pm \else $H^\pm$\fi}
\def\lum{\ifmmode {\cal L}\else ${\cal L}$\fi}
\def\inpb{\ifmmode {\rm pb}^{-1}\else ${\rm pb}^{-1}$\fi}
\def\infb{\ifmmode {\rm fb}^{-1}\else ${\rm fb}^{-1}$\fi}
\def\epem{\ifmmode e^+e^-\else $e^+e^-$\fi}
\def\ppb{\ifmmode \bar pp\else $\bar pp$\fi}

\newskip\zatskip \zatskip=0pt plus0pt minus0pt

\def\matth{\mathsurround=0pt}
\def\lsim{\mathrel{\mathpalette\atversim<}}
\def\gsim{\mathrel{\mathpalette\atversim>}}
\def\atversim#1#2{\lower0.7ex\vbox{\baselineskip\zatskip\lineskip\zatskip
  \lineskiplimit 0pt\ialign{$\matth#1\hfil##\hfil$\crcr#2\crcr\sim\crcr}}}

\renewcommand{\thefootnote}{\fnsymbol{footnote}}

\begin{document} \begin{titlepage}
\setcounter{page}{1}
\thispagestyle{empty}
\rightline{\vbox{\halign{&#\hfil\cr
&PURD-TH-93-13\cr
&December 1993\cr}}}
\vspace{0.1in}
\begin{center}
\vglue 2.0cm
{\Large\bf Precision Electroweak Tests \\}
\vspace{0.2cm}
{\Large\bf on the $Sp(6)_L \times U(1)_Y$ Model\\}
\vglue 1.5cm
{T.~K.~Kuo$^{(a)}$ and Gye~T.~Park$^{(b),(c)}$\\}
\vglue 1.5cm
{\em $^{(a)}$Department of Physics, Purdue University\\}
{\em West Lafayette, IN 47907, USA\\}
{\em $^{(b)}$Center for Theoretical Physics, Department of Physics, Texas A\&M
University\\}
{\em College Station, TX 77843--4242, USA\\}
{\em $^{(c)}$Astroparticle Physics Group, Houston Advanced Research Center
(HARC)\\}
{\em The Woodlands, TX 77381, USA\\}
\baselineskip=12pt

\end{center}

\begin{abstract}

We perform precision electroweak tests on the $Sp(6)_L \times U(1)_Y$ model.
The purpose of the analysis is to delineate the model parameters
such as the mixing angles of the extra gauge bosons present in this model.
We find that the model is already constrained considerably by the present
LEP data.

\end{abstract}

\renewcommand{\thefootnote}{\arabic{footnote}} \end{titlepage}
\setcounter{page}{1}


\section{Introduction}
Precision measurements at the LEP have been extremely successful in confirming
the validity of the Standard Model(SM)\cite{1inUV}. Indeed, in order to have
agreements between theory and experiments, one has to go beyond the tree-level
calculations and include known electroweak(EW) radiative corrections. However,
from the theoretical point of view, there is a concensus that the SM can only
be a low energy limit of a more complete theory. It is thus of the utmost
importance to try and push to the limit in finding possible deviations from the
SM. In fact, there are systematic programs for such precision tests. Possible
deviations from the SM can all be summarized into a few parameters which then
serve to measure the effects of new physics beyond the SM. A lot of efforts
have gone into this type of investigation trying to develop a scheme to
minimize the disadvantage of having unkown top quark mass$(m_t)$ but to
optimize sensitivity to new physics. To date significant constraint!
s have been placed on a number of
 the technicolor model\cite{RCTC}, and some extended gauge
models\cite{Altetal}.
In this work we wish to apply the analysis to another extension of the SM, the
$\sp6$ family model. Amongst several of the available parametrization schemes
in the literature, the most appropriate one for our purposes is that of
Altarelli et.~al\cite{ABJ}. This is because their $\epsilon$-parametrization
can be used for new physics which might appear at energy scales not far from
those of the SM. This is the case for the $\sp6$ model. We still find that
parameters in this model are severely constrained. Thus, the precision EW tests
have demonstrated clearly that they are powerful tools in shaping our searches
for
extensions of the SM.

In Sec.~II, we will describe the $\sp6$ model, spelling out in detail the parts
that are relevant to precision tests. In Sec.~III, we summarize properties of
the $\epsilon$-parameters which will be used in our analysis. Sec.~IV contains
our detailed numerical results. Finally, some concluding remarks are given in
Sec.~V.

\section{$\sp6$ Model}
The $SP(6)_L\otimes U(1)_Y$ model, proposed some time ago\cite{kuo-nakagawa},
is the simplest extension of the standard model of three generations
that unifies the standard $SU(2)_L$ with the horizontal gauge group
$G_H(=SU(3)_H)$ into an anomaly free, simple, Lie group. In this model,
the six left-handed quarks (or leptons) belong to a {\bf 6} of
$SP(6)_L$, while the right-handed fermions are all singlets. It is thus
a straightforward generalization of $SU(2)_L$ into $SP(6)_L$, with the three
doublets of $SU(2)_L$ coalescing into a sextet of $SP(6)_L$. Most of the
new gauge bosons are arranged to be heavy $(\geq 10^2$--$10^3\rm\,TeV)$ so as
to avoid sizable FCNC. $SP(6)_L$ can be naturally broken into $SU(2)_L$
through a chain of symmetry breakings. The breakdown
$SP(6)_L \rightarrow [SU(2)]^3 \rightarrow SU(2)_L$ can be induced by two
antisymmetric Higgs which tranform as $({\bf 1}, {\bf 14}, 0)$ under
$SU(3)_C\otimes SP(6)_L\otimes U(1)_Y$. The standard $SU(2)_L$ is to be
identified with the diagonal $SU(2)$ subgroup of
$[SU(2)]^3=SU(2)_1\otimes SU(2)_2\otimes SU(2)_3$, where $SU(2)_i$ operates
on the $i$th generation exclusively. In terms of the $SU(2)_i$ gauge boson
$\vec{A}_i$, the $SU(2)_L$ gauge bosons are given by $\vec{A}={1\over\sqrt 3}
(\vec{A}_1+\vec{A}_2+\vec{A}_3)$. Of the other orthogonal combinations of
$\vec{A}_i$,
$\vec{A}^\prime={1\over\sqrt 6}(\vec{A}_1+\vec{A}_2-2\vec{A}_3)$, which
exhibits unversality only
among the first two generations, can have a mass scale in the TeV range
\cite{1TeVZ}. The three gauge bosons $A^\prime$ will be denoted as $Z^\prime$
and $W^{\prime\pm}$.
Given these extra gauge bosons with mass in the TeV range, we can expect small
deviations from the SM. Some of these effects were already analyzed elsewhere.
For EW precision tests,
the dominant effects of new heavier gauge boson $Z^\prime (W^{\prime\pm})$ show
up
in its mixing with the standard $Z(W^\pm)$ to form the mass eigenstates
$Z_{1,2} (W_{1,2})$:
\[ \hbox to \hsize{$ \hfill
\begin{array}{rcl}
Z_1&=&Z\cos\phi_Z+Z^\prime\sin\phi_Z \;, \\
W_1&=&W\cos\phi_W+W^\prime\sin\phi_W \;,
\end{array} \quad
\begin{array}{rcl}
Z_2 &=& -Z\sin\phi_Z+Z^\prime\cos\phi_Z \;, \\
W_2 &=& -W\sin\phi_W+W^\prime\cos\phi_W \;,
\end{array} \hfill
\begin{array}{r}
\stepcounter{equation}(\theequation)\\
\stepcounter{equation}(\theequation)
\end{array}
$} \]
where $Z_1 (W_1)$ is identified with the physical $Z(W)$.
Here, the mixing angles $\phi_Z$ and $\phi_W$ are expected to be small
$(\lsim0.01)$, assuming that they scale as some powers of mass ratios.

With the additional gauge boson $Z^\prime$, the neutral-current Lagrangian
is generalized to contain an additional term
\begin{equation}
L_{NC}=g_Z J_Z^\mu Z_\mu +g_{Z^\prime} J_{Z^\prime}^\mu Z_\mu^\prime \;,
\end{equation}
where $g_{Z^\prime}=\sqrt{1-x_W\over 2} g_Z={g\over\sqrt{2}}$,
$x_W=\sin^2\theta _W$, and $g={e\over {\sin\theta _W}}$. The neutral currents
$J_Z$ and $J_{Z^\prime}$ are given by
\begin{eqnarray}
J_Z^\mu &=&\sum_{f} \bar{\psi}_f\gamma^\mu\left( g^f_V+g^f_A\gamma _5\right)
\psi_f \;, \\
J_{Z^\prime}^\mu &=&\sum_{f} \bar{\psi}_f\gamma^\mu\left( g^{\prime
f}_V+g^{\prime f}_A\gamma _5\right)
\psi_f \;,
\end{eqnarray}
where $g^f_V={1\over 2}\left( I_{3L}-2x_Wq\right)_f$, $g^f_A={1\over
2}\left( I_{3L}\right)_f$ as in SM, $g^{\prime f}_V=g^{\prime
f}_A={1\over 2}\left( I_{3L}\right)_f$
for the first two generations and $g^{\prime f}_V=g^{\prime
f}_A=-\left( I_{3L}\right)_f$ for the third. Here $\left(I_{3L}\right)_f$ and
$q_f$
are the third component of weak isospin and electric charge of fermion $f$,
respectively. And the neutral-current Lagrangian reads in terms of $Z_{1,2}$
\begin{equation}
L_{NC}=g_Z\sum_{i=1}^2\sum_{f} \bar{\psi}_f\gamma_\mu\left(
g^f_{Vi}+g^f_{Ai}\gamma _5\right)
\psi_f Z^\mu_i \;,
\end{equation}
where $g^f_{Vi}$ and $g^f_{Ai}$ are the vector and axial-vector
couplings of fermion $f$ to physical gauge boson $Z_i$, respectively.
They are given by
\begin{eqnarray}
g^f_{V1, A1}&=&g^f_{V, A}\cos\phi_Z+{g_{Z^\prime}\over g_Z} g^{\prime
f}_{V, A}\sin\phi_Z \;, \\
g^f_{V2, A2}&=&-g^f_{V, A}\sin\phi_Z+{g_{Z^\prime}\over g_Z} g^{\prime
f}_{V, A}\cos\phi_Z \;.
\end{eqnarray}
Similar analysis can be carried out in the charged sector.

\section{One-loop EW radiative corrections and the $\epsilon$-
parameters}
It is now well known that EW parameters become consistent with the data
only if the EW radiative corrections are accounted for.
For example,
the predictions for $\sin^2\theta_w$ and $M_W$, obtained from  various
 measurements at $M_Z$ and low-energy $\nu$ scattering experiments are
consistent only if
one-loop effects are included.

There are several different schemes to parametrize
the EW vacuum polarization corrections \cite{Kennedy,PT,efflagr,AB}. It
can be easily shown that by expanding the vacuum polarization tensors to order
$q^2$, one obtains three independent physical parameters. Alternatively, one
can
show that upon symmetry breaking there are three
additional terms in the effective lagrangian \cite{efflagr}.
In the $(S,T,U)$ scheme \cite{PT},
the deviations of the model predictions from those of the SM (with fixed
values of $m_t,m_H$) are considered to be  as the effects from ``new physics".
This scheme is
only valid to the lowest order in $q^2$, and is therefore not viable for a
theory with new, light $(\sim M_Z)$ particles. In the $\epsilon$-scheme, on the
other hand, the model predictions are absolute and are valid up to higher
orders in $q^2$, and therefore this scheme is better suited to
the EW precision tests of the MSSM\cite{BFC} and a class of supergravity models
\cite{PARKeps}.
Here we choose to use the $\epsilon$-scheme because the new particles in the
model to be considered here can be relatively light $(O(1TeV))$.
In this scheme,three independent physical parameters $\epsilon_{1,2,3}$
\cite{AB} correspond to a set of observables $\Gamma_{l}, A^{l}_{FB}$ and
$M_W/M_Z$.
Among these three
parameters, only $\epsilon_1$ provides very strong constraint
, for example, in
supersymmetric models\cite{ABCII,PARKeps}.
The expressions for $\epsilon_{1,2,3}$ are given as
\cite{BFC,PARKeps}
\begin{eqnarray}
\epsilon_1 &=& e_1-e_5-{\delta G_{V,B}\over G}-4\delta g_A\;,\\
\epsilon_2 &=& e_2-s^2 e_4-c^2 e_5-{\delta G_{V,B}\over G}
				-\delta g_V-3 \delta g_A\;,\\
\epsilon_3 &=& e_3+c^2 e_4-c^2 e_5+{c^2-s^2\over 2 s^2}\delta g_V
         -{1+2s^2\over 2 s^2}\delta g_A\;,
\end{eqnarray}
where $e_{1,..,5}$ are the following combinations of vacuum polarization
amplitudes
\begin{eqnarray}
e_1&=&{\alpha\over 4\pi \sin^2\theta_W M^2_W}[\Pi^{33}_T(0)-\Pi^{11}_T(0)]\;,\\
e_2&=&F_{WW}(M_W^2)-{\alpha\over 4\pi s^2}F_{33}(M_Z^2)\;,\\
e_3&=&{\alpha\over 4\pi s^2}[F_{3Q}(M_Z^2)-F_{33}(M_Z^2)]\;,\\
e_4&=&F_{\gamma\gamma}(0)-F_{\gamma\gamma}(M_Z^2)\;,\\
e_5&=& M_Z^2F^\prime_{ZZ}(M_Z^2)\;,
\end{eqnarray}
and the $q^2\not=0$ contributions $F_{ij}(q^2)$ are defined by
\beq
\Pi^{ij}_T(q^2)=\Pi^{ij}_T(0)+q^2F_{ij}(q^2).
\eeq
The quantities $\delta g_{V,A}$ are the contributions to the vector and
axial-vector form factors at $q^2=M^2_Z$ in the $Z\to l^+l^-$ vertex from
proper vertex diagrams and fermion self-energies, and $\delta G_{V,B}$ comes
from the one-loop box, vertex and fermion self-energy corrections
to the $\mu$-decay amplitude at zero external momentum. It is important to note
that these non-oblique corrections are non-negligible, and must be included
in order to obtain an accurate gauge-invariant prediction\cite{Lynnetal}.
However, we have included the Standard non-oblique corrections only, neglecting
justifiably the small effects from the new physics.
In the following section we
calculate $\epsilon_{1}$ in the
$\sp6$ model. We do not, however, include $\epsilon_{2,3}$ in our analysis
simply because these parameters can not provide any constraints at the current
level
of experimental accuracy\cite{BB,PARKeps}.
We assume throughout the analysis that the
non-oblique contributions from new physics to the measurables that are included
in the global fit are negligible.
Although loop corrections due to extra gauge bosons could be neglected
completely as in Ref\cite{Altetal}, we have improved the model prediction for
the oblique corrections by implementing the new vertices from Eq.~(6)
for the fermion loops only.
In this way we have accounted for a significant deviation of the model
prediction from the SM value for not so small $|\phi_{Z,W}|$.
Furthermore, in models with extra gauge bosons such as the model to be
considered here, the contribution from the mixings of these extra bosons with
the SM ones $(\Delta\rho_M)$ should also be added to
$\epsilon_1$\cite{Altetal,Altarelli90,parkkuo93}.

\section{Results and Discussion}

In order to calculate the model prediction for the Z width, it is sufficient
for our purposes to resort to the improved Born approximation (IBA)\cite{IBA},
neglecting small additional effects from the new physics.
Weak corrections can be effectively included
within the IBA, wherein the
vector couplings of all the fermions are determined by an effective
weak mixing angle.
In the case $f\not= b$, vertex corrections are negligible, and one
obtains the standard partial $Z$ width
\begin{equation}
\Gamma(Z\longrightarrow f\bar{f})=N^f_C\rho {G_FM^3_Z\over 6\pi\sqrt 2}\left(
1+{3\alpha\over 4\pi}q^2_f\right)\left[ \beta _f{\left( 3-\beta
^2_f\right)\over 2}{g^f_{V1}}^2+\beta^3_f {g^f_{A1}}^2\right] \;,
\end{equation}
where $N_C^f =1$ for leptons, and for quarks
\begin{eqnarray}
N_C^f &\cong&3\left[1+1.2{\alpha_S\left(
M_Z\right)\over\pi}-1.1{\left(\alpha_S\left(
M_Z\right)\over\pi\right)}^2-12.8{\left(\alpha_S\left(
M_Z\right)\over\pi\right)}^3\right] \;,\\
\beta_f&=&\sqrt {1-{4m_f^2\over M_Z^2}} \;, \\
\rho&=&1+\Delta\rho_M+\Delta\rho_{SB}+\Delta\rho_t\;, \\
\Delta\rho_t &\simeq& {3G_Fm_t^2\over 8\pi^2\sqrt 2} \;.
\end{eqnarray}
where the $\rho$ parameter includes not only the effects of the
symmetry breaking $\left(\Delta\rho_{SB}\right)$\cite{SB} and those
of the mixings between the SM bosons and the new bosons
$\left(\Delta\rho_{M}\right)$, but also the loop effects
$\left(\Delta\rho_{t}\right)$.
$N_C^f$ above is obtained by accounting for QCD corrections up to
3-loop order in $\overline{MS}$ scheme, and we ignore different QCD
corrections for vector and axial-vector couplings which are due not
only to chiral invariance broken by masses but also the large mass
splitting between $b$ and $t$. We use for the vector and axial vector couplings
$g^f_{V1}$ and $g^f_{A1}$ in Eq.~(7) the effective $\sin^2\theta_W$,
$\bar{x}_W=1-{M_W^2\over{\rho M_Z^2}}$.
In the case of $Z\longrightarrow b\bar{b}$,
the large $t$ vertex correction should be accounted for by the following
replacement
\begin{equation}
\rho \longrightarrow\rho-{4\over 3}\Delta\rho_t\,, \quad
\bar{x}_W\longrightarrow\bar{x}_W\left( 1+{2\over 3}\Delta\rho_t\right) \;.
\end{equation}

In the following analysis, we use the recent experimental value,
$\epsilon_1=(-0.9\pm3.7)\times 10^{-3}$, obtained from a global
fit to LEP data on $\Gamma_l,A^{l,b}_{FB},A^\tau_{pol}$ and $M_W/M_Z$
measurement\cite{BB,PARKeps}.
We consider not only a constraint on the deviation of $\Gamma_Z$ from the SM
prediction\cite{parkkuo93}, $\Delta\Gamma_Z\leq 14$~MeV, which is the present
experimental accuracy\cite{Luth}, but also the present experimental
bound on $\Delta\rho_M$.
We use a direct model-independent bound on $\Delta\rho_M$,
$\Delta\rho_M\lsim 0.0147-0.0043{\left({m_t\over 120 GeV}\right)}^2$ from
$1-({M_W\over{M_Z}})^2=0.2257\pm 0.0017$ and $M_Z=91.187\pm
0.007\GeV$\cite{Luth}.
The values $M_H=100\GeV$, $\alpha_S(M_Z)=0.118$, and $\alpha(M_Z)=1/128.87$
will be used
thoughout the numerical analysis.

In Fig.~1 we present the regions in the $(\phi_W, \phi_Z)$ plane excluded by
all the constraints to be imposed here for $m_t=130$, $M_{Z^\prime}=1000$, and
$M_{W^\prime}=800\GeV$. An excluded region shaded by horizontal lines
represents the $\epsilon_1$ constraint at $90\%$C.~L. whereas the one by
vertical lines corresponds to $\Delta\Gamma_Z$ and $\Delta\rho_M$ constraints.
We observe in Fig.~1 that $\epsilon_1$ starts cutting in the region
($\phi_Z\lsim-0.007$ and $\phi_W\gsim0.009$)
 still allowed by the other constraints even at $m_t=130\GeV$. Similarly in
Fig.~2 we also show the excluded regions for $m_t=170$, $M_{Z^\prime}=800$, and
$M_{W^\prime}=1000\GeV$.
It's very interesting for one to see two small disconnected allowed regions in
the figure. We find here that $\phi_Z\gsim 0$ and $\phi_W\gsim 0.005$ in one
region whereas $\phi_Z\gsim -0.0075$ and $\phi_W\gsim -0.002$ in the other.
The way that values for $M_{Z^\prime}$ and $M_{W^\prime}$ are chosen
originates from constraint due to $\Delta\rho_M$.
For $m_t=170\GeV$, the model predicts $\epsilon_1$ without $\Delta\rho_M$
too high to be allowed at $90\%$C.~L. for $|\phi_{Z,W}|\lsim0.01$. Therefore,
the model parameters including $M_{Z^\prime}$ and $M_{W^\prime}$ are chosen
in such a way that $\Delta\rho_M$ brings $\epsilon_1$ down to or below
the LEP bound at $90\%$C.~L. This is fulfilled only if
$M_{Z^\prime}<M_{W^\prime}$. However, for $m_t=130\GeV$,
$M_{Z^\prime}<M_{W^\prime}$ or $M_{Z^\prime}>M_{W^\prime}$ are allowed because
now $\epsilon_1$ lies always within the LEP bounds for
$|\phi_{Z,W}|\lsim0.01$.
For $M_{Z^\prime}>M_{W^\prime}$, $\Delta\rho_M$ brings $\epsilon_1$ up to
the LEP upper limit$(0.0052)$ at $90\%$C.~L. while for
$M_{Z^\prime}<M_{W^\prime}$
it brings $\epsilon_1$
down to the LEP lower limit$(-0.007)$. Since the contour lines for the one
choice are more or less those for the other choice with $90\deg$ rotation
around zero, we present in Fig.~1 only one choice, $M_{Z^\prime}>M_{W^\prime}$.
Moreover, for $m_t=170\GeV$, there are in fact two pairs of contour lines for
$\epsilon_1$ constraint. The pairs come from either the LEP upper limit or the
lower limit because $\epsilon_1$ can also go below the lower limit because of a
large negative constribution from $\Delta\rho_M$ for large mixing angles.

\section{Conclusions}

In this work we have concentrated on the constraints placed on the $\sp6$
family model from precision LEP measurements. As has been the cae with similar
studies, the model is severely constrained. The most important effects of the
model come from mixings of the SM gauge bosons $Z$ and $W$ with the additional
gauge bosons $Z^\prime$ and $W^\prime$. We have computed the one loop EW
radiative corrections due to the new bosons in terms of $\epsilon_1$ and
$\Delta\Gamma_Z$.
Using a global
fit to LEP data on $\Gamma_l,A^{l,b}_{FB},A^\tau_{pol}$ and $M_W/M_Z$
measurement, we find that the mixing angles $\phi_Z$ and $\phi_W$ are
constrained to lie in rather small regions. Also, larger ($\gsim 1\%$)
$\phi_Z$ and $\phi_W$ values are allowed only when there is considerable
cancellation between the $Z^\prime$ and $W^\prime$ contributions, corresponding
to $|\phi_Z|\approx|\phi_W|$. It is noteworthy that the results are sensitive
to the top quark mass.
For small $m_t$ $(130\GeV)$, the allowed parameter regions are considerably
bigger than those for larger $m_t$ $(170\GeV)$ values. Hopefully, when the top
quark mass becomes available, we can narrow down the mixing angles with
considerable precision.

\section*{Acknowledgements}

The work of T.K. has been supported in part by DOE. The work
of G.P. has been supported by a World Laboratory Fellowship.

\newpage

%
\def\NPB#1#2#3{Nucl. Phys. B {\bf#1} (19#2) #3}
\def\PLB#1#2#3{Phys. Lett. B {\bf#1} (19#2) #3}
\def\PLIBID#1#2#3{B {\bf#1} (19#2) #3}
\def\PRD#1#2#3{Phys. Rev. D {\bf#1} (19#2) #3}
\def\PRL#1#2#3{Phys. Rev. Lett. {\bf#1} (19#2) #3}
\def\PRT#1#2#3{Phys. Rep. {\bf#1} (19#2) #3}
\def\MODA#1#2#3{Mod. Phys. Lett. A {\bf#1} (19#2) #3}
\def\IJMP#1#2#3{Int. J. Mod. Phys. A {\bf#1} (19#2) #3}
\def\TAMU#1{Texas A \& M University preprint CTP-TAMU-#1}
\def\ARAA#1#2#3{Ann. Rev. Astron. Astrophys. {\bf#1} (19#2) #3}
\def\ARNP#1#2#3{Ann. Rev. Nucl. Part. Sci. {\bf#1} (19#2) #3}

\newpage

%
{\bf Figure Captions}
\begin{itemize}

\item Figure 1: The region excluded by the $\epsilon_1$ constraint at $90\%$
C.~L.(horizontal). The region excluded by $\Delta\Gamma_Z\leq 14 MeV$ and
$\Delta\rho_M$ constraint
(vertical).
$m_t=130\GeV$, $M_{Z^\prime}=1000\GeV$, and $M_{W^\prime}=800\GeV$ are used.
\item Figure 2: Same as in Figure.~1 except that $m_t=170\GeV$,
$M_{Z^\prime}=800\GeV$, and $M_{W^\prime}=1000\GeV$ are used.
\end{itemize}


\begin{thebibliography}{99}
\bibitem{SB} We assume that $\rho_{SB}=1$ even if it depends on a
particular Higgs structure.
\end{thebibliography}
\end{document}